\documentclass[prd,
nofootinbib,
preprintnumbers,
twocolumn,
showpacs
]{revtex4}
\usepackage{graphicx,epsf,epsfig}
\def\eq#1{{eq.~(\ref{#1})}}
\def\eqs#1#2{{eqs.~(\ref{#1})--(\ref{#2})}}

\def\vev#1{\left\langle #1\right\rangle}

\def\ie{{\it i.e. }}

\def\hbar{\hspace{0pt}\raisebox{1pt}{$-$} \hspace{-7pt} h}

\def\5{\overline 5}

\newcommand{\be}{\begin{equation}}
\newcommand{\ee}{\end{equation}}
\newcommand{\bd}{\begin{displaymath}}
\newcommand{\ed}{\end{displaymath}}
\newcommand{\bea}{\begin{eqnarray}}
\newcommand{\eea}{\end{eqnarray}}
\newcommand{\nn}{\nonumber}

\begin{document}

\preprint{UCRHEP-T377}
%\preprint{hep-ph/0406117}

%----------------------------------------------------------------------------------
\title{Fermion masses in SUSY SO(10) with type II seesaw:\\ a non-minimal predictive scenario}
%----------------------------------------------------------------------------------
\date{June 10, 2004}
\author{Stefano Bertolini $^{1}$}\email{bertolin@sissa.it}
\author{Michele Frigerio $^{2}$}\email{frigerio@physics.ucr.edu}
\author{Michal Malinsk\'{y} $^{1}$}\email{malinsky@sissa.it}
\affiliation{$^{1}$ Scuola Internazionale Superiore di Studi Avanzati,
via Beirut 4, I-34014 Trieste and INFN, Sezione di Trieste, Italy \\
$^{2}$Department of Physics, University of California, Riverside, CA 92521, USA}
%----------------------------------------------------------------------------------

\begin{abstract}

A predictive framework for fermion masses and mixing is given by the
supersymmetric $SO(10)$ model with one $10_{H}$, one $\overline{126}_{H}$,
one $126_{H}$ and one ${210}_{H}$ Higgs representations, and
type II seesaw dominating the neutrino mass matrix. We investigate the
origin of the tension between this model and lepton mixing data and refine
previous numerical analyses.
We discuss an extension of the minimal
model
that includes one $120_H$ chiral superfield representation.
This exhausts
the possible renormalizable contributions to the Yukawa sector.
In spite of the increase in the number of
parameters the predictivity of the minimal setting is not spoiled.
We argue that the contributions to fermion masses due
to the doublet components of $120_H$ can be
naturally small compared to those of $10_{H}$ and $\overline{126}_{H}$, thus
acting as a perturbation in the fermion mass generation.
The antisymmetric
nature of the $120_H$ Yukawa coupling affects
at leading order the determination of the mixing angles
and it allows to remove
the inconsistencies between predictions and data
on the neutrino parameters.
An improvement in the experimental bound on $U_{e3}$ can tell this scenario
from the minimal model.
\end{abstract}

%----------------------------------------------------------------------------------
%----------------------------------------------------------------------------------
\maketitle

%==========================================================================
\section{Introduction}
%==========================================================================

Remarkably enough only a few years after the last run of LEP, which has
marked the era of precision
electroweak laboratory tests, neutrino physics is reaching the age of
precision studies (for a recent
review see ref. \cite{Smirnov:2003xe}).
Neutrinos being massive, flavor mixing is present in the leptonic sector
as well, underlying,
together with the
known mass differences, the phenomenon of neutrino oscillations. The
smallness of their masses compared
to the other fermions and the evidence that the shape of the leptonic
mixings differs substantially from
the hierarchical structure of the quark mixings add a challenge in
understanding the origin of the fermionic spectrum.
Even more so when considering grand unified scenarios
where lepton
and quark Yukawa couplings are related by gauge symmetry.

Some interest has been revived in
the recent years on a supersymmetric (SUSY) implementation of $SO(10)$
which has a minimal number of
parameters~\cite{Aulakh:2003kg,Bajc:2004xe}
(as many as the Minimal Supersymmetric Standard Model (MSSM) with massive
neutrinos and exact R-parity) and
exhibits a remarkable predictivity
in the neutrino sector linking the maximality of the atmospheric mixing to
the $b-\tau$ Yukawa
unification \cite{Bajc:2002iw,Bajc:2001fe}.

The minimal renormalizable SUSY $SO(10)$ model \cite{Clark:ai,Aulakh:1982sw}
contains in addition to three generations
of $16_F$ matter supermultiplets the following Higgs chiral
supermultiplets:
$10_H$, $\overline{126}_H$, $126_H$, and $210_H$. The $10_H$ and
$\overline{126}_H$ representations
couple to the matter bilinear $16_{Fi} 16_{Fj} = (10_S + 126_S +
120_A)_{ij}$ in the
superpotential leading to the minimal set of Yukawa couplings needed for a
realistic fermion mass
spectrum \cite{Babu:1992ia} ($S,A$ denote the symmetry property in the generation indices
$i,j$). The $126_H$ representation
is needed in order to preserve supersymmetry from $D$-term breaking, while
the $210_{H}$ triggers the
$SO(10)$ gauge symmetry breaking and provides the needed mixings
among the Higgs supermultiplets.
Besides leading to a realistic matter mass spectrum, the model features
exact R-parity conservation, due
to the even congruency class ($B-L=2$) of the $10$ and  $126$
representations
($120$ shares the same property), with relevant implications for cosmology
and proton
decay \cite{Lee:1994je,Aulakh:1997fq,Aulakh:1999cd,Aulakh:2000sn,Bajc:2003ht,Goh:2003nv,Fukuyama:2004pb}.

The smallness of the neutrino masses naturally follows from the seesaw
mechanism which is present in a
twofold type:
\be
M_\nu = - M_{\nu_D}^T M_{\nu_R}^{-1} M_{\nu_D} + M_{\nu_L} ~.
\label{seesawrelation}
\ee
The first term represents the canonical (type I)
seesaw \cite{general-seesaw-type-I}. The Majorana mass matrix $M_{\nu_R}$
is generated by the vacuum expectation value (VEV)
of a $SU(2)_R$ triplet field in
$\overline{126}_H$, while $M_{\nu_D}$ is the Dirac neutrino mass matrix.
The second term (type II seesaw \cite{general-seesaw-type-II})
is present due to a very small VEV (proportional
to the square of the electroweak scale over the GUT scale) induced by the
weak scale breaking on the $SU(2)_L$ triplet component in
$\overline{126}_H$.

Assuming the dominance of type II seesaw one finds an intriguing
connection
~\cite{Bajc:2002iw,Bajc:2001fe} between $b-\tau$ unification
(which can be approximatively achieved even in the presence of
$\overline{126}_H$)
and the almost maximal atmospheric neutrino
mixing angle ($\sin^2 2\theta_{23}\gtrsim 0.9$ at $90\%$ C.L.
\cite{Nakaya:2002ki}).
On the other hand, the detailed numerical analysis which have been
carried out in refs. \cite{Mohapatra1,Mohapatra2}
show a number of possible
short-comings. First, the 1-3 lepton
mixing is bound to be quite close to the present experimental upper bound
($U_{e3}\lesssim 0.2$ at $90\%$ C.L. \cite{Apollonio:1999ae}).
Second, the solar mixing angle $\theta_{12}$ is predicted too close to
maximal, while the SNO result,
$\theta_{12}=32.5^{+2.4}_{-2.3}$ degrees \cite{Ahmed:2003kj},
definitely excludes this possibility.
Third, the deviation of atmospheric mixing $\theta_{23}$ from maximal is
too large.
When the effect of CP phases is taken into account, only the solar mixing
is significantly affected
and its fitted value can be in agreement with data, while $U_{e3}$ still
cannot be smaller
than $\approx 0.16$ \cite{Mohapatra2}.
On the other hand, the CKM phase is predicted to be in the second or third
quadrant requiring
significant contributions to CP violation from other sources \cite{Mohapatra3}.

Type I seesaw has been also investigated within the SUSY $SO(10)$
scenario \cite{Brahmachari:1997cq,Oda:1998na,Matsuda:2000zp}.
Even though one may recover in some limit the type II seesaw relation
between $b-\tau$ Yukawa unification and large neutrino atmospheric
mixing, a dominant type I seesaw contribution is highly
disfavored by the global fit of neutrino data, unless non-renormalizable
terms are added to the minimal model \cite{Mohapatra3,Bajc:2004fj}.

In the first part of this paper (Section \ref{minimalmodelestimate})
we present an independent study of the minimal renormalizable
SUSY $SO(10)$ model with dominant type II seesaw.
We work out simple analytic
arguments to explain the origin of the tension between the fermion mass
sum rules of the model
and the experimental values of the lepton mixing.
A new numerical fit is presented, including some
experimental uncertainties previously neglected.
We refine the results in
the literature, albeit confirming some of the shortcomings of the model in
reproducing the detailed structure of the neutrino parameters.

In order to improve the agreement with the data, one may certainly
consider
extensions of the minimal $SO(10)$ model, that include
additional $10_H$ and/or $\overline{126}_H$ representations.
However, the presence of the new set of Yukawa couplings spoils the
predictivity of the minimal model for fermion mass textures.
A more interesting option
is considering the presence of Planck-induced non-renormalizable
operators that make the $SO(10)$ model
an effective theory at the GUT scale. The presence of the new Yukawa terms
leads to some additional ($M_G/M_{Pl}$ suppressed) contributions
 in the fermion mass sum rules.
On the other hand, even though the size of the corrections is under
control,
the breakdown of renormalizability
allows for a number of possible effective scenarios whose effects
generally overlap, thus weakening again
the predictive feature of the minimal model.
In ref. \cite{Mohapatra3} the authors choose to consider effective
operators
that transform as $\overline{126}$ thus maintaining the symmetry
property of the renormalizable Yukawa terms.
Because of that the
fermion mass relations present strong similarities with the minimal model,
whose predictions remain to a large extent unmodified, while obtaining
some of the desired improvements.

In this paper we take the standpoint of maintaining renormalizability,
while considering the effects of adding to the minimal model content
a $120_H$ supermultiplet.
All possible renormalizable contributions
to the Yukawa sector are present (an early discussion can be found in
\cite{Barbieri:1979ag}).
We argue (Section \ref{theory}) that the induced VEVs of the bidoublet components of
$120_H$ can be naturally suppressed compared to the weak scale.
This allows for treating
the $120_H$ contributions to fermion masses as a perturbation,
thus preserving most of the features of the minimal model.
On the other hand, due to the different
symmetry properties of the new Yukawa couplings, we show that to leading
order in the fermion mass corrections an excellent fit
of the neutrino parameters is obtained. Future neutrino data may provide
a test of this scenario and discriminate it with respect to the
minimal one.

The analytic tools for the study of the $120_H$
contributions to the fermion mass textures are developed and discussed
in Section \ref{effectsof120}, while the numerical results and
the experimental signature of the extended renormalizable
$SO(10)$ model are presented in Section \ref{results}.

We limit the present analysis to real Yukawa coupling. This choice allows
to clearly evaluate
what is the weight of each quark (lepton) mass or mixing angle in the fit.
In fact, previous
studies \cite{Mohapatra2,Mohapatra3} show that the effect of the phases is
subleading and adds only
a minor freedom in reproducing the data.
Finally, the aim of the present discussion is to
emphasize the role of symmetry and size
of the $120_H$ corrections in predicting a realistic fermion mass spectrum.
The investigation of CP violation and possible connections between the CKM
phase and the CP phases in the lepton sector is left for future work.

%%%%%%%%%%%%%%%%%%%%%%%%%%%%%%%%%%%%%%%%%%%%%%%%%%%%%%%%%%%%%%%%%%%%%%%%%
\section{\label{minimalmodelestimate} Fermion masses and mixing in
the Minimal SUSY $SO(10)$ with type II seesaw}
%%%%%%%%%%%%%%%%%%%%%%%%%%%%%%%%%%%%%%%%%%%%%%%%%%%%%%%%%%%%%%%%%%%%%%%%%

As mentioned in the previous section, the Higgs sector
of the minimal renormalizable SUSY $SO(10)$ model
includes in addition to $10_H$,
the $\overline{126}_H$, $126_H$, and $210_H$ representations.
The $10_H$ and
$\overline{126}_H$ representations
couple to the matter bilinear $16_{F} 16_{F}$
in the
superpotential leading to the minimal set of Yukawa couplings needed for a
realistic fermion mass spectrum.
In this case the
fermion mass matrices are given by \cite{Babu:1992ia,Mohapatra:1979nn}
\bea
\label{minos} M_u &=& Y_{10}v_u^{10}+ Y_{126}v_u^{126}\nn ~,\\
M_d &=& Y_{10}v_d^{10}+ Y_{126}v_d^{126}\label{minimalrelations} ~,\\
M_l &=& Y_{10}v_d^{10}-3 Y_{126}v_d^{126}\nn ~,\\
M_\nu &\propto& Y_{126} ~,\nn
\eea
where dominant type II seesaw has been assumed for the neutrino mass matrix.
In general, the two light Higgs isodoublets $H_{u,d}$ are a linear
combination of the scalar isodoublets contained in $10$ and
$\overline{126}$ representations as well as of those contained in the
other Higgs representations not coupled to fermions. However, \eq{minos} is valid
independently on the composition of $H_{u,d}$, that does not affect the
following analysis. The only constraint is $\sum_i |v_i|^2=(174~{\rm GeV})^2$, 
where the sum runs over
all the isodoublets of the model.

The choice of type II seesaw is motivated by the connection between approximate $b-\tau$
unification at the GUT scale and almost maximal atmospheric mixing \cite{Bajc:2002iw,Bajc:2004fj}.
In fact,
\eq{minos} implies
\be
\label{releight} k \tilde{M_l}= \tilde{M_u} + r \tilde{M_d} \qquad \qquad M_\nu
\propto M_l - M_d ~,
\ee
where $\tilde{M_l}\equiv M_l/m_\tau$, $\tilde{M_u}\equiv M_u/m_t$ and
$\tilde{M_d}\equiv M_d/m_b$, while $k$ and $r$ are functions
of the VEVs in \eq{minos} and of $m_\tau,m_t,m_b$. Considering only the 2-3 blocks and neglecting
also second generation
masses, one can extract the relation between 2-3 quark and lepton mixings
\cite{Bajc:2002iw}: $$
\tan 2 \theta_{23}\approx \frac{2\sin \theta^q_{23}} {2 \sin^2\theta^q_{23}-\frac{m_b-m_\tau}{m_b}}
\qquad (\theta^q_{23}\approx 0.04)~. $$
Clearly, large atmospheric mixing requires cancellation between $m_b$ and $m_\tau$.
However, a complete three generation fit of fermion masses and mixing at the GUT scale is highly non-trivial.
In this paper we will limit our analysis to the case of no CP violation, where all mixing matrices and
mass eigenvalues are real.

%%%%%%%%%%%%%%%%%%%%%%%%%%%%%%%%%%%%%%%%%%%%%%%%%%%%%%%%%%%%%%%%%%%%%
\subsection{Understanding analytically the constraints on
the lepton mixing}
%%%%%%%%%%%%%%%%%%%%%%%%%%%%%%%%%%%%%%%%%%%%%%%%%%%%%%%%%%%%%%%%%%%%%

The standard approach \cite{Mohapatra1, Mohapatra2} to decipher
the predictions of this model is the following.
Let us rewrite \eq{releight} on the basis where $M_d$ is diagonal:
\bea
\tilde{M'_l} \equiv U_d^T\tilde{M_l}U_d & = &
\frac{1}{k} (V_{CKM}^T\tilde{D}_u V_{CKM} + r \tilde{D}_d)
~,\label{diagonalized} \\
M'_\nu\equiv U_d^T M_\nu U_d & = & m_0
\left[ \tilde{M'_l} - \frac{m_b}{m_\tau} \tilde{D}_d \right] ~,\label{nudia}
\eea
where $\tilde{D}_u={\rm diag}\left(\frac{m_u}{m_t},\frac{m_c}{m_t},1\right)$,
$\tilde{D}_d={\rm diag}\left(\frac{m_d}{m_b},\frac{m_s}{m_b},1\right)$,
$V_{CKM}$ is the Cabibbo-Kobayashi-Maskawa matrix
and $m_0$ is an overall neutrino mass scale.
All parameters are to be evaluated at the GUT
scale, since the relations in \eq{minos} are derived
at the scale where the unified gauge group
is broken to the Standard Model (SM).

The uncertainties in quark masses and mixing as well as the parameters
$k$ and $r$ can be used to fit the charged lepton masses via
eq. (\ref{diagonalized}). The equality of LHS and RHS traces implies
$k= 1+r+{\cal O}(\lambda^2)$ (where $\lambda\approx 0.22$ is
the Cabibbo angle).
It turns out (see e-print of ref. \cite{Mohapatra1})
that a better fit of the atmospheric mixing
is obtained for $m_s<0$ and $m_\mu>0$ (once third generation masses
are chosen positive). In this case, the requirement of
reproducing the correct value of $m_\mu/m_\tau$ leads to $k\approx 0.25$ and
$r\approx -0.75$.
Given $r$ and $k$ the RHS of eq. (\ref{nudia})
is completely determined and defines the neutrino mass matrix
up to an overall mass scale.

For the purpose of a simple analytical understanding of the
predictions in the lepton sector, we assume
for the time being $k=0.25$ and $r=-0.75$ exactly. Detailed variations in
$k$ and $r$ are taken into account in the numerical fit
discussed in section \ref{fitmin}. Using the  Wolfenstein
parametrization of $V_{CKM}$ and
neglecting ${\cal O}(\lambda^4)$ terms  we obtain
\begin{equation}
\begin{array}{l}
\tilde{M'_l} \equiv U_l \tilde{D_l} U_l^T =
\left(
\begin{array}{ccc} 0 & 0 & 4 V_{td} \\
\dots & -3\frac{m_s}{m_b} & 4 V_{ts} \\
\dots & \dots &  1
\end{array}
\right) ~,
\\
M'_\nu \equiv U_\nu D_\nu U_\nu^T = m_0
\left(
\begin{array}{ccc} 0 & 0 & 4 V_{td} \\
\dots & -\left(\frac{m_b}{m_\tau}+3\right)\frac{m_s}{m_b} & 4 V_{ts} \\
\dots & \dots &  1-\frac{m_b}{m_\tau}
\end{array}\right) ~,
\end{array}
\label{crucial}
\end{equation}
where we use the convention in which $V_{ts}$ and $m_s$ are
negative and $V_{td}$, $m_b$ and $m_\tau$ positive.
Notice that the ``parallel'' structure of $\tilde{M'_l}$ and
$M'_\nu$ is ``broken'' by $b-\tau$ unification.
It is impressive that the structure of these matrices reflects
qualitatively the basic features of lepton
masses and mixings: $\tilde{M'_l}$ is hierarchical with small
mixing angles and
$$
\frac{m_\mu}{m_\tau}\approx -3 \frac{m_s}{m_b} -
16 V_{ts}^2 ~.
$$
Both large 1-2 and 2-3 mixing should be contained in $M'_\nu$:
this is the case since
the elements in the 2-3 block of $M'_\nu$
can be taken of the same order (dominant $\mu\tau$-block) and
the 1-3 element is automatically smaller. As a consequence
\cite{Frigerio:2002rd}, the spectrum of neutrinos
is predicted to be with normal hierarchy.

In ref.~\cite{Bajc:2004fj} the exact computation of the leptonic 2-3
mixing is performed for the present model in the case of two generations.
The authors find two solutions for large mixing: one corresponds
to the scenario described
above: $\tilde{M'_l}$ hierarchical and $b-\tau$ unification inducing
large 2-3 mixing in $M'_\nu$. The
second solution corresponds to $r\approx -1$ (and $|k|\ll 1$),
leading to a cancellation in the $33$-entry of
$\tilde{M'_l}$. However, it can be easily shown that this possibility
is spoiled by a three generation analysis. In fact, we find that in this
case the charged lepton mass matrix has the form
\begin{equation}\label{Mlforrminus1}
\tilde{M'_l}|_{r\approx -1}=-\frac{m_b}{m_s a}
\left(
\begin{array}{ccc} 0 & 0 & V_{td} \\
\dots & -\frac{m_s}{m_b} & V_{ts} \\
\dots & \dots &  -\frac{V^2_{ts}m_b}{m_s}
\end{array}\right) +{\cal O}(\lambda^4) ~,
\end{equation}
where $a\approx (1+V^2_{ts}m^2_b/m^2_s)$ is of order unity.
This structure can be suitable to generate a small $m_\mu/m_\tau$
in the case of two generations, but since the determinant of
$\tilde{M'_l}$ is of order $\lambda^2$, it is clear that the
hierarchy $m_e\div m_\mu\div m_\tau\approx \lambda^5\div \lambda^2\div 1$
cannot be reproduced.

Let us analyze in some detail eq.~(\ref{crucial}).
The matrices $\tilde{M'_l}$ and $M'_\nu$ depend only on four quark
parameters ($V_{td},~V_{ts},~m_s/m_b,~m_b/m_\tau$) and they are required to
reproduce five lepton parameters ($m_\mu/m_\tau$, $\Delta m^2_{\odot}/\Delta
m^2_{A}$, $\theta_{12}$, $\theta_{13}$ and $\theta_{23}$), where $\theta_{ij}$
are the mixing angles in the lepton mixing
matrix $U_{PMNS}\equiv U_l^T U_\nu$.
The first generation masses $m_e$ and $m_1$ are sensitive also
to subleading terms neglected in eq. (\ref{crucial}).
Since the quark parameters are known with small uncertainties,
there is very small freedom to fit lepton data.
Notice that, in good approximation, we can compare directly the neutrino
masses and mixing angles obtained from eq.~(\ref{crucial}) at the GUT scale with
the experimental values at the electroweak scale. As a matter of fact, in the case of normal
hierarchy, the RGE running of the neutrino mass matrix has a negligible
effect on mass squared differences and mixings
\cite{Casas:1999tg,Frigerio:2002in,Antusch:2003kp}.

For the mixing angles in $\tilde{M'_l}$ and $M'_\nu$ we use the notation
$c^{l,\nu}_{ij}\equiv \cos\theta^{l,\nu}_{ij}$ and
$s^{l,\nu}_{ij}\equiv \sin\theta^{l,\nu}_{ij}$.
Since $4 m_s/m_b\sim 4 V_{ts}\lesssim \lambda$, the 2-3 mixing in
$M'_\nu$ is generically of order unity if $b-\tau$ unification
is realized to $\lambda\div\lambda^2$ accuracy.
In general, the deviation from maximal mixing increases
with $\Delta m^2_{\odot}/\Delta m^2_{A}$.
The other two mixing angles in $M'_\nu$ are given approximately by
\be
\label{bound}
\begin{array}{c} s_{13}^\nu \approx (c^\nu_{23})^3
\frac{4V_{td}}{1-\frac{m_b}{m_\tau}}
= {\cal O}(\lambda)~,\\
\sin 2\theta^\nu_{12} \approx s_{23}^\nu (c^\nu_{23})^2
\frac{8V_{td}}{1-\frac{m_b}{m_\tau}}
\sqrt{\frac{\Delta m^2_{A}}{\Delta m^2_{\odot}}}
= {\cal O}(1) ~,
\end{array}
\ee
where we used
$\sqrt{\Delta m^2_{\odot}/\Delta m^2_{A}}\sim \lambda$.
If one neglects the small mixing in $\tilde{M'_l}$, all
oscillation data can be reproduced.
For example, taking $m_b/m_\tau =0.89$, $V_{ts}=-0.035$, $V_{td}=0.011$,
$m_s/m_b=-0.028$, one obtains $s^\nu_{13}\approx 0.12$,
$\tan\theta^\nu_{23}\approx 0.97$, $\tan^2\theta^\nu_{12}\approx 0.43$ and
$\Delta m^2_{\odot}/\Delta m^2_{A}\approx 0.038$.

However, it turns out that the small mixing angles in ${\tilde M'_l}$
add up to those in $M'_\nu$ in such a way to
spoil the agreement with data.
In fact, we find
\begin{equation}
\begin{array}{c}
\theta_{23}^l\approx 4V_{ts}\approx -0.14\sim -\lambda ~,~~~
\theta_{13}^l\approx 4V_{td}\approx 0.04\sim\lambda^2 ~,\\
\theta_{12}^l\approx -\frac{16V_{ts}V_{td}}{3m_\mu/m_\tau}
\approx 0.10\sim\lambda ~,
\end{array}
\end{equation}
where the numerical estimates are obtained from the input values at the end of
the previous paragraph. The effect of the two $O(\lambda)$ rotations in
$U_l$ modifies the physical mixing angles in
$U_{PMNS}\equiv U_l^T U_\nu$ as follows:
\begin{equation}
\begin{array}{c}
\theta_{23}\approx \theta_{23}^\nu+\theta_{23}^l ~,~~~ s_{13}\approx s_{13}^\nu+s_{12}^ls_{23}^\nu ~,\\
\sin2\theta_{12}\approx\sin2\theta_{12}^\nu
\left(1+s_{12}^l\frac{2c_{23}^\nu}{\tan2\theta_{12}^\nu}\right) ~.
\end{array}
\end{equation}
As a consequence, to reproduce data one needs $\theta_{23}^\nu$
larger than $\pi/4$, $s_{13}^\nu$ significantly below the experimental
upper bound and $\sin 2\theta_{12}^\nu$ smaller than the solar mixing value.
Both the deviation from $\theta^\nu_{23}=\pi/4$ and the suppression of
$\sin2\theta_{12}^\nu$
tend to increase the ratio $\Delta m^2_{\odot}/\Delta m^2_{A}$
above the allowed range, producing a tension between predictions and experimental data, as confirmed by
the numerical study that follows.

%%%%%%%%%%%%%%%%%%%%%%%%%%%%%%%%%%%%%%%%%%%%%%%%%%%%%%%%%%%%%%%%%%%%%%%%%%
\subsection{\label{fitmin} Discussion of the  numerical results}
%%%%%%%%%%%%%%%%%%%%%%%%%%%%%%%%%%%%%%%%%%%%%%%%%%%%%%%%%%%%%%%%%%%%%%%%%%

Previous numerical analysis of fermion masses and mixing in SUSY SO(10) with $10$ and $\overline{126}$ Higgs
fields coupled to matter
and type-II seesaw dominance are
given in refs. \cite{Mohapatra1,Mohapatra2}.
All studies find a tension between lepton mixings and quark
parameters: $s_{13}$ turns out to be close to the present upper bound
($\approx 0.16$), the atmospheric mixing can be hardly  close enough to maximal ($\sin^2 2
\theta_{23}\lesssim 0.9$) and the solar mixing is too large to fit the LMA MSW solution ($\sin^2
2\theta_{12}\gtrsim 0.9$). The last
drawback can be relaxed tuning CP violating phases, but in disagreement with the known value of the CKM
phase
\cite{Mohapatra2,Mohapatra3}.

We have run an independent fit of the experimental data paying
particular attention to the uncertainties in the input parameters.
It is in fact crucial to determine how far the minimal $SO(10)$ scenario
can be pushed in reproducing the known fermion spectrum and mixings.
Due to the complexity of the numerical analysis,
we have taken advantage of
the analytical results derived in the previous section
to elaborate an efficient approach to the fit,
while obtaining a rationale for the emerging patterns.

We input
the GUT-scale values of quark masses given in ref.~\cite{Das:2000uk}
for two typical values of $\tan \beta$, namely
$\tan \beta=10$ and $\tan \beta=55$,
and consider both 1- and 2-$\sigma$ ranges.

Our numerical fit confirms the results of ref.~\cite{Mohapatra2}
for the central values of the quark mixing angles and $\Delta
m^{2}_{\odot}/\Delta m^{2}_{A}\lesssim 0.05$ there considered. In
this case we find $\sin^{2}2\theta_{23}\lesssim 0.93$,
$|U_{e3}|\approx 0.16$ and $\sin^2
2\theta_{12} \gtrsim  0.92 $, the latter beeing excluded at the $90\%$ C.L..
 When we include the 1-$\sigma $ uncertainties in the $V_{CKM}$
entries \cite{Hagiwara:fs} and allow for $0.019\le \Delta
m^{2}_{\odot}/\Delta m^{2}_{A} \le 0.069$ (the 90\% C.L.
experimental range \cite{Nakaya:2002ki,Eguchi:2002dm,Ahmed:2003kj}) we do not find any
major deviation due to the $V_{CKM}$ entries, the largest effects being related to the 
extended $\Delta m^{2}_{\odot}/\Delta m^{2}_{A}$ range. 
For $\tan\beta=10$ larger
values (albeit not maximal) of the atmospheric neutrino mixing
angle are allowed, the upper bound being $\sin^2
2\theta_{23}\lesssim 0.97$, together with a reduced solar mixing, namely 
$\sin^2 2\theta_{12} \gtrsim  0.85$ (the extreme values are obtained for $\Delta m^{2}_{\odot}/\Delta
m^{2}_{A}$ close to the 90\% C.L. upper bound).
The predictions for $|U_{e3}|$ is not significantly modified:
$|U_{e3}|\approx 0.16$ % for large atmospheric angle
.
For $\tan \beta=55$ the results are quite similar: the upper bound for the
atmospheric mixing is reduced to $\sin^2 2\theta_{23}\lesssim 0.95$, while the solar angle lower bound is
relaxed to $\sin^2 2\theta_{12} \gtrsim 0.82$. The $|U_{e3}|$ parameter is bound to be about $0.15$.

Only when the scan is performed over the 2-$\sigma$ ranges of the quark sector parameters,
maximal atmospheric mixing is allowed, while the lower bound for $|U_{e3}|$ can be reduced to about
$0.14$ and the solar mixing angle can be lowered to 0.75 (such a values are reached
for large $\tan\beta$ and for $\Delta m^{2}_{\odot}/\Delta m^{2}_{A}$ close to the
$90\%$ C.L. upper bound).

As pointed out in ref. \cite{Bajc:2004fj}, the two neutrino
analysis suggests the possible relevance of the parameter region
characterized by $r\approx -1$
(corresponding to the solution $\sigma=+1$ in the notation of ref.
\cite{Bajc:2004fj}),
where the atmospheric mixing may be {\it naturally} large.
However, we have checked that in this domain one can not recover a
good fit of the electron mass,
in full agreement with our argument after \eq{Mlforrminus1}.
In Sect. V we display some graphics of the results here discussed
compared to those of the extended model introduced in the next section.

In conclusion our numerical analysis confirms the patterns found
by previous authors and analytically discussed in the previous
subsection. The minimal renormalizable SUSY $SO(10)$ model,
while providing a suggestive and  appealing framework for understanding the
main features of the quark and lepton spectra, fails in reproducing
the data at the present 1-$\sigma$ experimental accuracy. When considering
the 2-$\sigma$ experimental ranges, agreement with the data is obtained
in limited regions of the parameter space.
This motivated us to study the effects of extending the minimal
model to include one $120_H$ representation.

%==================================================================================
\section{$120_H$-extension of the minimal renormalizable SUSY
SO(10) \label{theory}}
%==================================================================================
In this section we discuss how
the inclusion of one $120_H$ representation
in the minimal renormalizable SUSY-SO(10)
may affect the electroweak breaking pattern.
Using a simplified and self-explanatory notation
the superpotential of the extended model
reads:
\bea
W_Y & = & 16_F \left( Y_{10} 10_H +Y_{126} \overline{126}_H +Y_{120} 120_H
\right)16_F \nn \\
W_H & = & M_{10} 10_{H}^2+M_{126} \overline{126}_{H} 126_{H} +M_{210} 210_{H}^2 + \nn \\
& + & M_{120} 120_H^2 +\lambda\, 210_{H}^3+\eta\,210_{H}\overline{126}_{H}126_{H}+ \nn \\
& + & 10_{H} 210_{H} (\alpha\,126_{H}+\beta\,\overline{126}_{H})+ \nn \\
& + &  \eta'\,210_{H}120_{H}120_{H}+ \gamma\,10_{H}210_{H}120_{H}+ \nn \\
& + & 120_{H}210_{H}(\alpha'\,126_{H}+ \beta'\,\overline{126}_{H})
\label{superpot}
\eea
where the $3\times 3$ matrices (in general complex) $Y_{10}$ and $Y_{126}$
are symmetric while $Y_{120}$ is antisymmetric.

We do not report in detail the Higgs potential derived from \eq{superpot},
the D-terms and the scalar soft breaking terms.
Since $120_H$ does not contribute to $SO(10)$ breaking, we
may assume that the correct breaking
pattern to the minimal supersymmetric standard model (MSSM) is
achieved~\cite{Aulakh:2003kg,Bajc:2004xe}.
$SO(10)$ can be broken spontaneously down to the MSSM either directly or via
one or more intermediate steps~\cite{Rajpoot:xy,Mohapatraetal}.
For the discussion that follows it is convenient to write down the
explicit decomposition of the
$SO(10)$ Higgs representations under the Pati-Salam (PS) subgroup
$SU(4)_{PS}\otimes SU(2)_L\otimes SU(2)_R \equiv G_{422}$:
\bea
10 & = & (1,2,2)\oplus (6,1,1)\nn \\
\overline{126} & = & (15,2,2)\oplus (\overline{10},3,1) \oplus (10,1,3)
\oplus (6,1,1)\nn \\
120 & = & (1,2,2)\oplus (15,2,2)\oplus (\overline{10},1,1) \oplus (10,1,1)
\nn \\
& &\oplus (6,3,1)\oplus (6,1,3)\nn \\
210 & = & (15,1,3)\oplus (15,3,1)\oplus (\overline{10},2,2) \oplus
(10,2,2) \nn \\
& & \oplus (6,2,2)\oplus (15,1,1)\oplus (1,1,1)
\label{Higgsrep}
\eea
It is also helpful to recall the $SU(3)_c\otimes U(1)_{B-L}$
decomposition of the $SU(4)_{PS}$
multiplets (with standard model $B-L$ normalization):
\bea
6 & = & (3,-2/3)\oplus (\overline{3},+2/3) \nn \\
10 & = & (6,+2/3)\oplus (3,-2/3)\oplus (1,-2) \nn \\
15 & = & (8,0)\oplus (3,+4/3)\oplus  (\overline{3},-4/3)\oplus(1,0)
\label{su3b-ldecomp}
\eea
%----------------------------------------------------------------------------------
%\subsection{Symmetry breaking and tree-level VEVs}
%----------------------------------------------------------------------------------

A non vanishing VEV of the
$(1,1,1)_{210}$ ($(15,1,1)_{210}$) component of $210_H$ triggers the
breaking
of $SO(10)$ down to $G_{422}$ ($G_{3221}$). We denote the scale of this
spontaneous breaking by $M_G$.
The subsequent left-right (LR) symmetry
breaking step to the SM group $G_{321}$ is achieved at
the scale
$M_R \leq M_G$ by VEVs of $(15,1,3)_{210}$, $(\overline{10},1,3)_{126}$
and $(10,1,3)_{\overline{126}}$.
Since the $B-L$ charge of the color singlets contained in $10$
(and $15$) of $SU(4)_{PS}$ is even, R-parity is
preserved.
The study of gauge coupling unification in SUSY $SO(10)$
favours the scenario of the direct $SO(10)\to SU(3)_c\otimes SU(2)_L\otimes
U(1)_Y \to SU(3)_c\otimes U(1)_Q$ breaking chain
(see, e.g., ref.~\cite{Bajc:2004xe}).
As a consequence we will henceforth take $M_R \approx M_G \approx 10^{16}$ GeV.

The final electroweak breaking step is obtained by the VEVs induced by
weak scale SUSY soft-breaking terms on the light LH doublets obtained
from the
colorless components of the bidoublets
present in \eq{Higgsrep}.
Since $210_H$ mixes $126_H$, as well as $120_H$, with $10_H$ one expects
that all the color (and $B-L$) singlet LH doublets
mix to give (via fine tuning)
the two light Higgs doublet superfields of the MSSM, leaving the other
states heavy.

In this respect the bidoublet components in the $120_H$ representation
may exhibit a specific feature.
Since no $120_H$ component participates to the spontaneous
breaking of $SO(10)$ down to the MSSM group
we may consider the value of its mass parameter $M_{120}$
in \eq{superpot} as naturally taken at the cut-off scale of the model
i.e. the Planck mass $M_{Pl}$ \cite{delAguila:at}.
As a consequence, one expects decoupling
effects
of $120_H$ proportional to $M_G/M_{Pl}\approx 10^{-3}$.
In particular the $120_H$ colorless $SU(2)_{L}$ doublet components acquire
an induced VEV suppressed
by the above factor with respect to the doublets contained in the other
representations.

Relations among the VEVs of the relevant components
can be obtained, neglecting explicit soft SUSY breaking,
from the F-term (and D-term) flatness
of the supersymmetric vacuum, i.e. by requiring
$\vev{F_X}=\vev{\partial W /\partial X}=0$
for any superfield $X$ in the superpotential,
replaced by its scalar component.

Considering the $SO(10)$ superpotential in \eq{superpot}
and its decomposition in terms of $G_{422}$, of which
some relevant terms are
\bea
126_{H}\overline{126}_{H}210_{H} & = &
(\overline{10},1,3)_{126}
(10,1,3)_{\overline{126}} (15,1,1)_{210} \nn \\
& + &  \ ...  \nn \\
10_{H}120_{H}210_{H} & = & (15,2,2)_{120}(1,2,2)_{10}(15,1,1)_{210} \nn \\
& + &  (1,2,2)_{120}(1,2,2)_{10}(1,1,1)_{210} \ + \ ... \nn \\
10_{H}126_{H} 210_{H} & = & (15,2,2)_{126}(1,2,2)_{10}(15,1,1)_{210} \nn
\\
& + & (10,3,1)_{126}(1,2,2)_{10}(\overline{10},2,2)_{210} \nn \\
& +&  ... \nn \\
10_{H}\overline{126}_{H} 210_{H} & = &  (10,1,3)_{\overline{126}}(1,2,2)_{10}(\overline{10},2,2)_{210} \nn \\
& +&  ... \ , \nn
\eea
the vacuum F-flatness in the $120_H$ bidoublet directions yields
\bea
\label{inducedVEVs120}
\langle(15,2,2)_{120}\rangle
& \sim &
\frac{M_{R}^2}{M_{120}M_{210}}
\langle(1,2,2)_{10}\rangle
\nn \\
\langle(1,2,2)_{120}\rangle
 &\sim &
\frac{M_{G}}{M_{120}}\langle(1,2,2)_{10}\rangle  \ ,
\eea
where
$O(1)$ couplings are assumed
and
$\vev{(\overline{10},1,3)_{126}} = \vev{(10,1,3)_{\overline{126}}} \sim M_R$
(as required by D-flatness at $M_R$).

In an analogous way for the colorless $\overline{126}_H$ LH components
one obtains
\bea
\label{inducedVEVs126}
\langle(15,2,2)_{\overline{126}}\rangle % & \sim &
& \sim & \frac{M_{R}^2}{M_{126}M_{210}}\langle(1,2,2)_{10}\rangle
\nn \\
\langle(\overline{10},3,1)_{\overline{126}}\rangle % & \sim &
& \sim &
\frac{M_{R}}{M_{210}}\frac{\langle(1,2,2)_{10}\rangle^2}{M_{126}}  \ ,
\eea
Notice in \eq{inducedVEVs126}
the very small VEV induced on the $\overline{126}_H$ LH
triplet by the weak breaking, leading
in the SUSY case ~\cite{Aulakh:1997fq,Aulakh:1999cd,Melfo:2003xi}
to the type II seesaw term
in \eq{seesawrelation}.

Considering the one-step breaking of $SO(10)$ ($M_R\sim M_G$)
from \eqs{inducedVEVs120}{inducedVEVs126}
 and the assumption $M_{120} \sim M_{Pl}$
one obtains that the $120_H$ LH doublet VEVs
are suppressed by $O(M_G/M_{Pl})$ with respect to those in
$10_H$ and $\overline{126}_H$. Since this result is controlled
by the decoupling of the $120_H$ representation,
the suppression of the $120_H$ VEVs
is not spoiled by the soft SUSY breaking potential which triggers
at the weak scale the $SU(2)_L \times U(1)_Y$ breaking.
After the needed minimal fine tuning,
the two light Higgs doublets
have $120_H$ components suppressed by $O(M_G/M_{Pl})$
so that on the broken vacuum \eq{inducedVEVs120} is reproduced.
This feature represents the basic ingredient in the following
discussion of fermion masses and mixings.

The role of $120_H$ can be replaced by Planck-scale induced
non-renormalizable operators which transform accordingly.
On the other hand, ad hoc assumptions
on the ultraviolet completion of the model and on the symmetry properties
of the effective couplings are generally needed in order to reproduce
the minimal (renormalizable) setup here discussed.

%%%%%%%%%%%%%%%%%%%%%%%%%%%%%%%%%%%%%%%%%%%%%%%%%%%%%%%%%%%%%%%%%%%%%%%%%%%%%%%
\section{\label{effectsof120} $120_H$-corrections to fermion masses and mixing}
%%%%%%%%%%%%%%%%%%%%%%%%%%%%%%%%%%%%%%%%%%%%%%%%%%%%%%%%%%%%%%%%%%%%%%%%%%%%%%%

The most general structure of the fermion mass matrices in the renormalizable $SO(10)$ model with all possible
types of Higgs fields coupled to fermions
is given by \cite{Barbieri:1979ag}
\begin{eqnarray}
\label{relations} M_u &=& Y_{10}v_u^{10}+ Y_{126}v_u^{126}+ Y_{120} v_u^{120} ~,\nn \\
M_d &=& Y_{10}v_d^{10}+ Y_{126}v_d^{126}+ Y_{120} v_d^{120} ~,\nn \\
M_l &=& Y_{10}v_d^{10}-3 Y_{126}v_d^{126}+ Y_{120} v_l^{120} ~,\nn\\
M_\nu &\propto& Y_{126} ~,
\end{eqnarray}
where the VEVs $v^{120}_{x}$ are three independent linear combinations of the four
$120_H$ isodoublet VEVs, and type II seesaw is assumed to dominate in $M_\nu$.
Motivated by the discussion in the previous section,
we take $v_x^{120}/v_x^{10,126} \sim M_G/M_{Pl}$, such that the $120_H$
contributions to the mass matrices can be treated as a small perturbation.

The analogue of eq. (\ref{releight}) now reads
\bea
k \tilde{M_l} & = & \tilde{M_u} + r \tilde{M_d}+Y_{120}(k \varepsilon_l-\varepsilon_u-r \varepsilon_d)
~,\nn \\
\label{simplified1}
M_\nu & \propto & \left[M_l - M_d+Y_{120}(m_b \varepsilon_d-m_\tau \varepsilon_l)\right] ~,
\eea
with the short-hand notation
$$
\varepsilon_u\equiv \frac{v_u^{120}}{m_t} ~,\qquad
\varepsilon_d\equiv \frac{v_d^{120}}{m_b} ~,\qquad
\varepsilon_l\equiv \frac{v_l^{120}}{m_\tau} ~.
$$
The mass matrices of charged fermions are asymmetric and can be diagonalized by means of a
biorthogonal transformation ($M_x=V^R_x D_x {V^L_x}^T,~x=u,d,l$), so that
\be
\label{Mlgeneral} k {V^R_d}^T\tilde{M_l}V^L_d= W^T\tilde{D}_u V_{CKM} + r \tilde{D}_d + Y_{120}'
(k \varepsilon_l-\varepsilon_u-r \varepsilon_d)
\ee
where
\be\label{various}
W\equiv{V^R_u}^T V_d^R ~,~~
V_{CKM}\equiv {V_u^L}^T V_d^L ~,~~
Y_{120}'\equiv{V^R_d}^T Y_{120}V^L_d ~.
\ee

The missing ingredient needed
to perform the fitting procedure of charged lepton masses, in analogy to the
minimal model case, is the right-handed quark mixing matrix $W$.
Since for $\varepsilon_x=0$ one has $W=V_{CKM}$, it is convenient to write
$W$ as $V_{CKM}$ plus order $\varepsilon_{x}$ corrections.
One obtains (see Appendix \ref{AppW})
\be
\label{w} W=V_{CKM}+2\left(-\varepsilon_u Z'_u V_{CKM}+ \varepsilon_d V_{CKM} Z'_d\right) +
{\cal O}(\varepsilon_x^{2}) ~,
\ee
where the antisymmetric matrices $Z'_{u,d}$ are given by
\be
\label{Zdprime} (Z'_{x})_{ij}=\frac{(Y_x')_{ij}}{(\tilde{D}_x)_{ii}+(\tilde{D}_x)_{jj}}
\ee
and
$Y'_{u} \equiv  V_{CKM} Y_{120}' V_{CKM}^T$, $Y'_{d} \equiv Y_{120}'$.

As shown in Appendix \ref{AppW}, the antisymmetry of $Y_{120}$
implies that the eigenvalues of the symmetric mass matrices
are unmodified up to ${\cal O}(\varepsilon_x^2)$ corrections.
This suggests that the $120_H$ induced mass corrections
may affect at the leading order the determination of the mixing angle
in such a way not to destabilize the good fit of the
mass eigenvalues obtained in the minimal model.
This feature is relevant for understanding qualitatively
the numerical discussion presented in the next section (see also Appendix~B).

The type II neutrino mass matrix now reads
\be
\begin{array}{c}
{V^L_d}^T{M_\nu}V^L_d =m_0 {V_d^L}^T V_d^R \left[{V^R_d}^T\tilde{M_l}V^L_d - \right. \\
\left.-\frac{m_b}{m_\tau}
\tilde{D}_d + Y_{120}'\left(\frac{m_b}{m_\tau}
\varepsilon_d - \varepsilon_l\right)\right] ~,
\label{Mnugeneral}
\end{array}\ee
where $m_0$ is an overall neutrino mass scale.
Using ${V_d^L}^T V_d^R \approx 1+2\varepsilon_d Z'_d$ and ${V_d^L}^T V_u^R =
V_{CKM}^T (1+2\varepsilon_u Z'_u)$
(see Appendix~\ref{AppW}), as well as eq. (\ref{Zdprime}),
one obtains
\bea\label{munu}
{V^L_d}^T{M_\nu}V^L_d  & =  &
M'_\nu+\Delta M'_\nu ~,
\eea
where $M'_\nu$ is given in \eq{nudia} and
\be
\begin{array}{c}
\Delta M'_\nu =  \frac{m_0}{k}
\left[\varepsilon_u V^T_{CKM}\left(2 Z'_u\tilde{D}_u-Y'_u\right) V_{CKM} +\right.\\
\left.+\varepsilon_d\left(r-\frac{m_b}{m_\tau}k\right)\left(2 Z'_d\tilde{D}_d-Y'_d\right) \right] ~.
\end{array}
\ee
Using eq. (\ref{Zdprime}) and taking into account the hierarchy
among quark masses,
one obtains, up to order $\varepsilon_x$ corrections ($x=u,d$),
\be
\left(2 Z'_x\tilde{D}_x-Y'_x\right)_{ij} \approx (Y'_{x})_{ij} sign(j-i) \equiv (Y^s_{x})_{ij}~.
\ee
Neglecting for simplicity ${\cal O}(\lambda)$ terms one can write $V_{CKM}^{T}Y^s_{u}V_{CKM}\sim
Y^s_{d}\equiv Y^s_{120}$ which finally leads to
\be \label{DeltaM}
\Delta M'_\nu \approx \frac{m_0}{k}\left[\varepsilon_u+\varepsilon_d \left(r-\frac{m_b}{m_\tau}k\right)
\right] Y^s_{120} ~.
\ee
The form of $\Delta M'_\nu$ allows  for a direct and simple assessment
of the effects of the $Y'_{120}$-matrix
entries on the minimal model neutrino mass spectrum and lepton mixings.
It is interesting that
$\Delta M'_\nu$ does not depend on $\varepsilon_{l}$ and therefore originates
entirely from the quark sector corrections.

Once we have reconstructed the matrices on the left-hand side of \eq{Mlgeneral} and \eq{munu},
the lepton mixing matrix $U_{PMNS}$ is given by
\be
\label{UPMNS} U_{PMNS}\equiv {U_l}^T U_\nu ~,%= U_1^T U_2
\ee
where
\bea
{V_d^L}^T M_l^T M_l V_d^L \equiv U_l D_l^2 U_l^T ~,~~
{V_d^L}^TM_\nu V_d^L \equiv U_\nu D_\nu U_\nu^T ~.\nn
\eea
Notice that \eq{UPMNS} does not depend on $V^L_d$.

%==================================================================================
\section{Numerical study of SUSY $SO(10)$ with $120_H$-corrections \label{results}}
%==================================================================================

We are now ready to present the results of the numerical analysis
that accounts for the effects of the $Y_{120}$ contributions on the
fermion mass matrices.
Together with the GUT-scale quark mass ranges and
mixings given in ref.
\cite{Das:2000uk} (for $\tan \beta=10,\ 55$),
we need to input the following additional set of parameters:
$Y_{120}'$, $\varepsilon_l$, $\varepsilon_u$, $\varepsilon_d $.
For simplicity in the present discussion all
CP-phases are set to zero.

\begin{figure}[ht]
\caption{\label{solatm} $\sin^2 2\theta_{12}$ as a function of $\sin^2 2\theta_{23}$ in the minimal
model (black) and for the $120_H$-extension with the set of parameters specified in the text (gray).
The dot-dashed contour encloses the experimentally allowed region at the
$90\%$ C.L.}
\epsfig{file=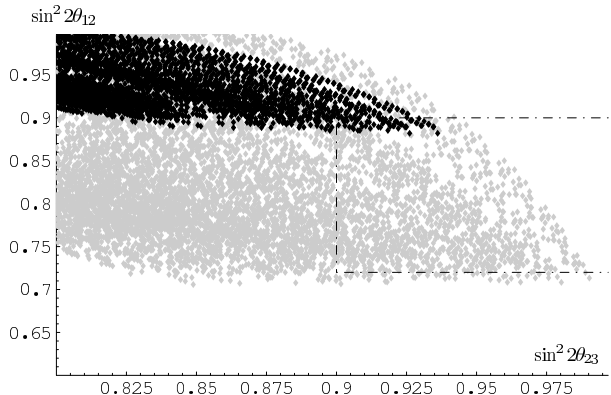 ,width=8cm}
\end{figure}

\begin{figure}[ht]
\caption{\label{deltam} $\Delta m^{2}_{\odot}/\Delta m^{2}_{A}$ as a function of $\sin^2 2\theta_{23}$
in the minimal model
(black) and for the $120_H$-extension with the set of parameters specified in the text (gray).
The dot-dashed contour encloses the experimentally allowed region at $90\%$ C.L.}
\epsfig{file=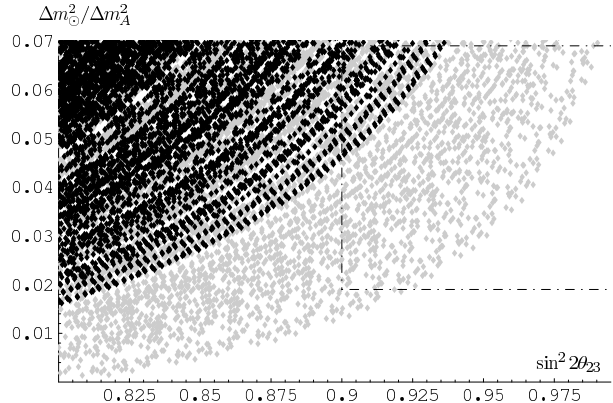 ,width=8cm}
\end{figure}

We perform an extensive scan within the allowed quark mass and
mixing ranges. For each point within the scanned region,
$W$ is given by eq. (\ref{w}). Using this input, we search for
values of $r$ and $k$ such that the charged lepton masses obtained from
\eq{Mlgeneral}
fit the charged-lepton data.
For each $r$ there remains the
freedom to shift $m_{b}$ (and/or  $m_{t}$)
together with $m_{d},m_{s}$,$v^{120}_{d}$
($m_{u},m_{c}$,$v^{120}_{u}$)
within the allowed ranges,
while keeping $\tilde{D}_{u,d}$ and $\varepsilon_{u,d}$ constant.
For different values of $m_{b}$,
one set of parameters fitting \eq{Mlgeneral} is
mapped into another fitting set
with different solutions of
the neutrino
mass matrices in eq.~(\ref{Mnugeneral}).
This procedure generates as a numerical artifact the
'chains' of solutions that are visible in Figs. \ref{solatm} and \ref{deltam}.

For illustration purposes we present our results for
$\varepsilon_l=0$ ($\Delta M'_{\nu}$ does not depend on $\varepsilon_{l}$) and for the following form of
the antisymmetric matrix $Y_{120}'$:
\be
\label{Y120here} Y_{120}'= a
\left(
\begin{array}{ccc} 0 & 1 & 1 \\
-1 & 0 & -1 \\
-1 & 1 & 0
\end{array}
\right)~.
\ee
As shown in Appendix \ref{fitexplanation}, thanks to the reduced values of
$\Delta m^2_\odot/\Delta m^2_A$ that are obtained, the texture in
\eq{Y120here} allows
for a substantial suppression
of the solar mixing angle with respect
to the corresponding minimal model solutions ($a=0$) as well as for reduced values of $|U_{e3}|$.

The parameters $v^{120}_{u,d}$ are given by
$$
v^{120}_{u,d} \sim \frac{M_G}{M_{Pl}}v^{10}_{u,d}\approx
10^{-1}(\sin\beta,\cos\beta)\ {\rm GeV}.
$$
Since $(Z'_{u,d})_{12}\approx (Y'_{120})_{12} m_{t,b}/m_{c,s}$
(see eq. (\ref{Zdprime})), the
expansion in eq.~(\ref{w}) of $W$ to leading order in $\varepsilon_{u,d}$
turns out to be  a good approximation
for $a\approx 0.1$.
The typical size of the $\varepsilon_{u,d}{Y}^{\prime,s}_{120}$ terms in eqs. (\ref{Mlgeneral}) and
 (\ref{DeltaM})  evaluated at the GUT scale and for
$\tan \beta=10$ is then given by
\bea
\varepsilon_u ({Y}^{\prime,s}_{120})_{ij} & \approx & a \frac{10^{-1} {\rm GeV}}{m_t}
\approx {\cal O}(10^{-4}) ~,\nn \\
\varepsilon_d ({Y}^{\prime,s}_{120})_{ij} & \approx & a \frac{10^{-2} {\rm GeV}}{m_b}
\approx {\cal O}(10^{-3}) ~.\label{setup}
\eea

\begin{figure}[ht]
\caption{\label{Ue3} $|U_{e3}|$ as a function of $\sin^2 2\theta_{23}$
in the minimal model (in black) and
for the $120_H$-extension with the set of parameters specified
in the text (in gray).
The dot-dashed contour encloses the experimentally allowed region at
the $90\%$ C.L.}
\epsfig{file=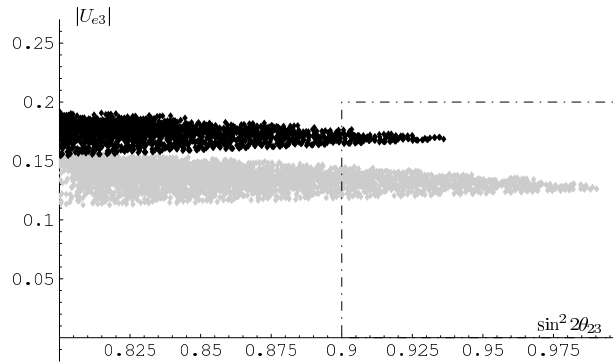 ,width=8cm}
\end{figure}\vspace{0.3cm}

To compare in an effective way the deviations obtained in the extended $SO(10)$
scenario with respect to the minimal model
results,
we present some of relevant allowed parameter planes
for $\tan\beta=10$ and 1-$\sigma$ ranges of
the quark masses, while taking the central values of the quark mixing angles.
Considering the $90\%$ C.L. range $0.019\leq \Delta m^2_\odot/\Delta m^2_A \leq 0.069$,
the allowed
area for the $U_{e3}$ parameter as a function of $\sin^2 2 \theta_{23} > 0.8$ is shown in Fig.~\ref{Ue3}.
The minimal model value $|U_{e3}|\approx 0.16$ is reduced by
the $120_H$ corrections to $0.11 < |U_{e3}| < 0.14$.
Even within such constrained setup
the atmospheric mixing is allowed to
be well within the 90\% C.L. experimental region
($0.90 \leq \sin^2 2\theta_{23} \leq 1$)
and in fact can be close to maximal.

In Fig.~\ref{solatm} we illustrate the change of the predicted values of
$\sin^2 2\theta_{12}$ as a function of $\sin^2 2\theta_{23}$.
The lower bounds on $\sin^2 2\theta_{12}$, which can be clearly seen both for the minimal model
and for the $120_H$-extension, are determined by the $90\%$ C.L. experimental upper bound
$\Delta m^2_\odot/\Delta m^2_A \leq 0.069$ that we have here considered.
In the extended model one obtains $\sin^2 2\theta_{12} \gtrsim 0.71$, thus covering the
whole $90\%$ C.L allowed range, while in the minimal model $\sin^{2}2\theta_{12}\gtrsim 0.88$.
For $\Delta m^2_\odot/\Delta m^2_A \lesssim 0.05$ one obtains
$\sin^2 2\theta_{12} \gtrsim 0.92$ and $0.77$ for the minimal and extended models respectively.
The presence of the $120_H$-corrections allows,
by reducing the values of $\Delta m^2_\odot/\Delta m^2_A$
(see Appendix B),
for lower (larger) values of the solar (atmospheric) angle.

The atmospheric mixing angle as a function of $\Delta m^2_\odot/\Delta m^2_A$
is shown in Fig.~\ref{deltam}.
Notice that, for the central value of the mass squared ratio ($\Delta m^2_\odot/\Delta m^2_A \approx
0.035$),
a significant
deviation from maximal atmospheric mixing
is present ($\sin^2 2\theta_{23} \lesssim 0.96$) in the extended model.

%==================================================================================
\section{Discussion and conclusions}
%==================================================================================

In this paper we have investigated the predictions of the renormalizable
supersymmetric SO(10) GUT for the masses and mixing angles of quarks and leptons.
Only two symmetric Yukawa coupling matrices, $Y_{10}$ and $Y_{126}$, determine
$M_u$, $M_d$, $M_l$ and $M_\nu$ in the minimal model.
We assumed that the low energy neutrino mass matrix is generated via type II seesaw, so that $M_\nu\propto Y_{126}$.
The model gives insight on the physics of flavor, the rationale being the
following:
there exists a weak basis
in which $Y_{10}$ is almost diagonal with hierarchical eigenvalues and dominates
the charged fermion mass matrices; at the same time $Y_{126}$ contains two large mixings
that show up in neutrino oscillations and provides the subleading corrections necessary
to explain the CKM mixing and the differences among $M_u$, $M_d$ and $M_l$ eigenvalue ratios.
The simplicity of this framework suggests that such weak basis could be identified
with the basis in which a flavor symmetry is realized.

We have reanalyzed the predictions of the minimal SUSY $SO(10)$
model for the leptonic sector, once the
quark data are considered.
The neutrino mass matrix has a dominant 2-3 block,
which implies that: i) neutrino spectrum has normal hierarchy;
ii) the contribution of neutrino
masses to neutrinoless $2\beta$ decay rate is negligible compared to
near future experimental sensitivity.
We have shown analytically that the 2-3 mixing is generically of order
unity,
but that, in order to keep the ratio of $Y_{126}$ eigenvalues
($\approx \sqrt{\Delta m^2_\odot/\Delta m^2_{A}}$) small,
a deviation from maximal mixing
of order $\lambda$ is required, which is inconsistent with atmospheric data.
A similar problem (but due to different constraints among the model parameters)
affects the 1-2 mixing:
$\sin 2\theta_{12} \propto (\Delta m^2_\odot/\Delta m^2_A)^{-1/2}$
and the suppression of the mass ratio leads to a too large mixing (order $\lambda$ above
the experimental value).
Finally, the 1-3 mixing is enhanced with respect to the quark sector by the
approximate $b-\tau$ unification and by a sizable ($\sim \lambda^2$) 1-3 entry in the
charged lepton sector: one finds $|U_{e3}|\approx 0.16$ for large $\theta_{23}$.
When complex phases are considered, a partially destructive interference between the neutrino and
charged lepton contribution to $\theta_{12}$ can be realized
\cite{Mohapatra2}, thus reconciling the model with solar data. However the other shortcomings of the
minimal model fit (namely $|U_{e3}|$ close to the upper bound and large deviation from maximal 2-3 mixing)
persist.

In summary,
including 1-$\sigma$ uncertainties in quark masses and mixings and taking $\tan\beta$ in the interval
$10-55$,
the predicted ranges for the
lepton mixing
angles in the minimal model (for $\Delta m^2_\odot/\Delta m^2_{A}\lesssim 0.05$) are:
$\sin^{2}2\theta_{12}\gtrsim 0.89$, $\sin^{2}2\theta_{23}\lesssim 0.97$,
$|U_{e3}|\gtrsim 0.15$.
Considering 2-$\sigma$ uncertainties in the quark input data
and the $90\%$ C.L. range for $\Delta m^2_\odot/\Delta m^2_{A}$,
the solar angle can be
as low as $\sin^{2}2\theta_{12}\gtrsim 0.75$
and $\theta_{23}$ is allowed to be maximal
(for large $\tan \beta$ and $\Delta m^2_\odot/\Delta m^2_{A}$ close to the upper
bound), while $0.14\lesssim|U_{e3}| \lesssim 0.16$.

In spite of the apparent tension with neutrino data,
we still find remarkable that the gross features
of fermion masses and mixing are reproduced within the minimal
renormalizable framework. This suggests that
small perturbations of the mass matrices
may be relevant to fitting the data.
We have considered a renormalizable extension of the minimal model
which includes
the antisymmetric component of $16_F\otimes 16_F$, that is, the $120_H$
representation.
We argued that the VEVs induced by soft SUSY breaking
on the $120_H$ Higgs doublet components can be
naturally suppressed by $O(M_G/M_{Pl})$ with respect to those of $10_H$ and
$\overline{126}_H$, thus preserving to large extent
the minimal model predictivity.
In addition, we showed that the asymmetry of the $120_H$ Yukawa coupling
plays a key role in fitting the flavor mixing data.

We developed a perturbation method to describe the small asymmetry induced in the quark
mass matrices and to estimate its effect on the charged lepton mixing. We found that
the $120_H$ corrections to the neutrino mass matrix affect to leading order
only the off-diagonal entries.
The 1-2 mixing in $U_{PMNS}$ and $\Delta m^2_\odot/\Delta m^2_A$ are more sensitive
to the correction because, in the case of normal hierarchy, these parameters are related to small quantities in
the neutrino mass matrix, which is dominated by  the atmospheric sector.
We showed how the $120_H$ contribution can be used to decrease the predicted mass squared ratio thus
relaxing the phenomenological problems of the minimal model.

Even assuming for simplicity zero CP violating phases,
we have shown that this framework allows already at 1-$\sigma$ level
for a consistent fit of all present data on fermion masses and mixing.
In particular, it is possible to reproduce values of
$\sin^2 2\theta_{23} \approx 1$, $\sin^2 2\theta_{12} \approx 0.8$ and $0.11\lesssim |U_{e3}|\lesssim
0.15$,
 in complete agreement with the current observed leptonic mixing data.

A positive evidence of $|U_{e3}| < 0.15$,
that is within the reach of forthcoming experiments, can reject
the minimal renormalizable $SO(10)$ model
and quantify the size of the $120_H$ correction.
We find that, for natural choices of the $120_H$ VEVs and couplings, a lower bound $|U_{e3}|\gtrsim 0.11$
is expected.
Moreover, at 1-$\sigma$ a deviation from maximal 2-3 mixing is expected
($\sin^2 2\theta_{23}\lesssim
0.97$), unless $\Delta m^2_\odot/\Delta m^2_{A}$ is large (close to the present upper bound).

\subsection*{Note added}

After the completion of this work, a more precise determination of the solar mass 
squared difference became available thanks to the release of a 
new  data set from the 
KamLAND experiment~\cite{0406039:2004mb}, that gives 
$\Delta m^2_\odot = (8.2^{+0.8}_{-0.6}) 10^{-5}$ eV$^2$ 
 at $90\%$ C.L. (raising the best fit value by about 20\%).
An updated combined analysis of SuperKamiokande and K2K data on 
atmospheric neutrinos~\cite{Maltoni:2004ei} implies
$\Delta m^2_A = (2.3^{+0.7}_{-0.9}) 10^{-3}$ eV$^2$ at $90\%$ C.L..
The mass squared ratio allowed range at 90\% C.L. is thus given by
$0.025\lesssim \Delta m^2_\odot / \Delta m^2_A \lesssim 0.064$ with best 
fit value $\approx 0.036$.
Notice that the uncertainty is still dominated by the atmospheric mass squared difference.
The reduced upper bound further restricts the corner 
in the parameter space where the minimal model may be viable (see 
Fig.~\ref{deltam}). The predicted lower bound on $\sin^2 2\theta_{12}$ becomes larger 
and therefore the tension with the smaller preferred experimental value 
increases (see Fig.~\ref{solatm}). 
These considerations strengthen the case for the $120_H$ extension of 
the model here proposed.

%%%%%%%%%%%%%%%%%%%%%%%%
\acknowledgments
%%%%%%%%%%%%%%%%%%%%%%%%%

M.F. thanks SISSA for hospitality and INFN for financial support during the early developments of this
project and K.S.~Babu and R.N.~Mohapatra for useful discussions.
The work of M.F. has been partially supported by the U.S. Department of Energy under Grant No.
DE-FG03-94ER40837.

%%%%%%%%%%%%%%%%%%%%%%%%%
\appendix
%%%%%%%%%%%%%%%%%%%%%%%%%

%%%%%%%%%%%%%%%%%%%%%%%%%%%%%%%%%%%%%%%%%%%%%%%%%%%%%%%%%%%%%%%%%%%%%%%%%%%%%%%%%%%%%%%%%%%%%%%%%%%%%%%
\section{Antisymmetric perturbation to a symmetric matrix \label{AppW}}
% Here we put the derivation of the relations presented in the text for W, Z, Z', Y'120 etc.
%%%%%%%%%%%%%%%%%%%%%%%%%%%%%%%%%%%%%%%%%%%%%%%%%%%%%%%%%%%%%%%%%%%%%%%%%%%%%%%%%%%%%%%%%%%%%%%%%%%%%%%

Consider a real symmetric matrix $S$
normalized so that the magnitude of its largest eigenvalue is 1.
There exists an
orthogonal matrix $U$ such that $S=U S^d U^T$ where $S^d$ is  diagonal.
If one adds a (real) antisymmetric matrix $\varepsilon A$ with
 $|A_{ij}|\leq 1$ and $\varepsilon\ll 1$,  a pair of orthogonal
matrices can be found such that $S+\varepsilon
A=V_1(\varepsilon) X^d(\varepsilon) V_2(\varepsilon)^T$.
Up to ${\cal
O}(\varepsilon^2)$ terms one gets
$$
V_1(\varepsilon)  =  (1+\varepsilon Z)U ~,~~
V_2(\varepsilon)  =  (1-\varepsilon Z)U ~,~~
X^d(\varepsilon)  =  S^d ~,
$$
where the antisymmetric matrix $Z$ satisfies
$$
\{S^d,U^T Z U \}=U^T A U ~.
$$
Denoting $U^T Z U \equiv Z'$ and $U^T A U \equiv A'$, one obtains
$$
Z'_{ij}=\frac{A'_{ij}}{S^d_{ii}+S^d_{jj}} ~.
$$
Proof: from $(S+\varepsilon A)^T=S-\varepsilon A$ we get
$V_1(-\varepsilon)=V_2(\varepsilon)$ and
$\quad X^d(-\varepsilon)=X^d(\varepsilon)$ which yields $X^d(\varepsilon)=S^d+{\cal O}(\varepsilon^2)$.
Expanding now
$S+\varepsilon A=V_1(\varepsilon) X^d(\varepsilon) V_1(-\varepsilon)^T$ with the ansatz
$V_1(\varepsilon)\equiv (1+\varepsilon Z)U$ (where $Z$ is antisymmetric by orthogonality of $V_1$)
one obtains, to the leading order in $\varepsilon$, $A=\{Z,S\}$. The last step is to rewrite this relation
in the diagonal basis for $S$.

These results allow us
to estimate the form of the right-handed quark mixing matrix $W$ in the presence of $120_H$-perturbation.
The quark mass matrices in eq. (\ref{relations}) can be written as
$$%\be
\tilde{M}_u = \frac{1}{m_t}M_u^s+\varepsilon_u Y_{120} ~, \qquad
\tilde{M}_d = \frac{1}{m_b}M_d^s+\varepsilon_d Y_{120} ~.
$$%\ee
Here $M_{u,d}^s$ are  the minimal model symmetric mass matrices, \ie the pieces $Y_{10}v_{u,d}^{10}+
Y_{126}v_{u,d}^{126}$. If the antisymmetric pieces $\varepsilon_i Y_{120}$ are very
small compared to the symmetric part, the eigenvalues of $M_{u,d}^s$  coincide with
those of the full $M_{u,d}$ up to ${\cal O}(\varepsilon^2)$ terms (while such corrections can be
relevant for first generation masses they are negligible for the estimate of mixing angles).
This implies, up to ${\cal O}(\varepsilon^2)$ terms,
$$
\tilde{M}_{x}= U_x
\tilde{D}_{x}{U_x}^T +\varepsilon_x Y_{120} = V^R_x
\tilde{D}_{x} {V^L_x}^T ~,
$$
for $x=u,d$. The orthogonal matrices $V_{x}^{R,L}$ are given by
$$
V^L_{x} =  (1-\varepsilon_{x} Z_{x})U_{x} ~,~~~
V^R_{x} =  (1+\varepsilon_{x} Z_{x})U_{x} ~.
$$
and the antisymmetric $Z_{x}$ satisfy
$$
\{\tilde{D}_{x},U_{x}^T Z_{x} U_{x}\}= U_{x}^T Y_{120} U_{x} ~.
$$
Using eq. (\ref{various}) and $Z'_{x}\equiv U_x^T Z_x U_x$, one obtains
\bea
W & = & (1-\varepsilon_u Z'_u){U_u}^T{U_d} (1+\varepsilon_d Z'_d) ~,\nn \\
V_{CKM}& =& (1+\varepsilon_u Z'_u){U_u}^T{U_d} (1-\varepsilon_d Z'_d) ~.\nn
\eea
This proves
eq. (\ref{w}).

%%%%%%%%%%%%%%%%%%%%%%%%%%%%%%%%%%%%%%%%%%%%%%%%%%%%%%%%%%%%%%%%%%%%%%%%%%%%%%%%%%%%%%%%%%%%%%%%%%%%%%%
\section{Example of $120_H$-effect on neutrino parameters \label{fitexplanation}}
%%%%%%%%%%%%%%%%%%%%%%%%%%%%%%%%%%%%%%%%%%%%%%%%%%%%%%%%%%%%%%%%%%%%%%%%%%%%%%%%%%%%%%%%%%%%%%%%%%%%%%%

In the minimal model without $120_H$-contribution,
the possibility of increasing the atmospheric
mixing and decreasing the solar one to get into the allowed region is prevented by the upper bound on
$\Delta m^2_\odot/\Delta m^2_A$.
As shown in Section \ref{results}, this problem can be removed in our scenario.
We give a simple analytical argument to prove that the $120_H$-correction to the neutrino mass
matrix, eq.~(\ref{DeltaM}), can be used to decrease the predicted value of
$\Delta m^2_\odot/\Delta m^2_A$.

The minimal model neutrino mass matrix $M'_\nu$ in eq.~(\ref{crucial}) can be written as
\be
\label{Mparametrisation}  M_\nu' = m_0\lambda
\left(
\begin{array}{ccc} X \lambda^2 & Y \lambda^3 & D \lambda \\
\dots & A & C \\
\dots & \dots & B
\end{array}
\right)~,
\ee
where $X,Y,A,B,C,D$ are ${\cal O}(1)$ parameters.
This texture generates the following neutrino spectrum hierarchy: $m_{1}\div m_{2}\div m_{3}$
$\sim \lambda \div\lambda \div 1$, the sign of $m_2$ being opposite to that of $m_3$ and $m_1$.
Assuming the setup defined by eqs.~(\ref{Y120here}) and (\ref{setup}),
we can estimate the leading contribution to $\Delta M'_\nu$ using eq. (\ref{DeltaM}):
$$
\Delta M'_\nu \approx -\frac{m_0}{k}\varepsilon_d Y^s_{120}\sim - m_0\lambda^3
\left(
\begin{array}{ccc} 0 & 1 & 1 \\
\dots & 0 & -1 \\
\dots & \dots & 0
\end{array}
\right) ~.
$$
The three independent quantities
${\rm Tr}M$, ${\rm Tr}M^2$ and ${\rm det}M$ characterize completely
the spectrum of a generic 3x3 real symmetric matrix $M$.
Using the parametrization (\ref{Mparametrisation}) one obtains
\bea
{\rm Tr}M'_\nu &=  & m_0 \lambda [A+B+X \lambda^2] ~,
\nn \\
{\rm Tr}(M'_\nu)^{2} & = & m_0^2 \lambda^2 [A^2+B^2+2C^2+  \nn \\
&& + 2 D^2\lambda^2 + X^2 \lambda^4 + 2 Y^2\lambda^6 ] ~,%
\nn \\
{\rm det}M_{\nu}' & = & m_0^3\lambda^3 [XAB \lambda^2 - A D^2\lambda^2 -  \nn \\
\label{abc} & &  - C^2 X\lambda^2 + 2 DCY\lambda^4 -B Y^2\lambda^6]~.
\eea
The addition of  $\Delta M'_\nu$ corresponds to the replacements
$Y \to  Y - {\cal O}(\lambda^{-1}) $,
$D  \to  D - {\cal O}(\lambda)$  and
$C  \to  C + {\cal O}(\lambda^2)$, so that
$$\begin{array}{l}
\delta {\rm Tr}M'_\nu=0 ~,\\
\delta {\rm Tr}(M'_\nu)^{2}\approx
m_0^2\lambda^2(4C\lambda^2)<0 ~,\\
\delta {\rm det}M'_\nu \approx m_0^3\lambda^3\cdot 2D\lambda^3(A-C)>0 ~.
\end{array}$$
Therefore $\delta(m_1+m_2+m_3)=0$, $\delta(m_1^2+m_2^2+m_3^2) \sim -m_0^2\lambda^4 $,
and $\delta(m_1m_2m_3) \sim m_0^3 \lambda^6$.
By writing the neutrino masses $m_i$ as the sum of the minimal model value $m_{i}^{0}$ plus the
$120_H$-correction $\delta_{i}$, one obtains
\bea
\delta_3(m^0_3-m^0_1)+ \delta_2(m^0_2-m^0_1)& \sim  & -m_0^2 \lambda^4 ~, \nn \\
\delta_3 m^0_2(m^0_1-m^0_3)+ \delta_2 m^0_3(m^0_1-m^0_2)& \sim  & + m_0^3 \lambda^6 ~.  \nn
\eea
which after some algebra yields
$\delta_2/m_{2}^{0} \sim - \lambda$
and $\delta_{3}/m_{3}^{0}\sim -\lambda^{2}$.
The ratio of mass squared differences is shifted as follows:
\bea
\frac{\Delta m^2_\odot}{\Delta m^2_A}\to
\frac{\Delta m^2_\odot}{\Delta m^2_A}
\left[1+2\frac{\delta_2}{m_2^0}+\ldots
\right].
\eea
Therefore the predicted value
of $\Delta m^2_\odot/\Delta m^2_A$ is reduced with respect to the
minimal model by a factor $\sim (1-2\lambda)$.


\begin{thebibliography}{99.}
%-------------------------------------------------------------------------------------------------
%\cite{Smirnov:2003xe}
\bibitem{Smirnov:2003xe}
A.~Y.~Smirnov,
%``Neutrino physics: Open theoretical questions,''
Int.\ J.\ Mod.\ Phys.\ A {\bf 19} (2004) 1180
[arXiv:hep-ph/0311259];
%%CITATION = HEP-PH 0311259;%%
%-------------------------------------------------------------------------------------------------
%\cite{Smirnov:2004ju}
%\bibitem{Smirnov:2004ju}
A.~Y.~Smirnov,
%``Neutrinos: '...Annus mirabilis',''
arXiv:hep-ph/0402264.
%%CITATION = HEP-PH 0402264;%%
%-------------------------------------------------------------------------------------------------
%\cite{Aulakh:2003kg}
\bibitem{Aulakh:2003kg}
C.~S.~Aulakh, B.~Bajc, A.~Melfo, G.~Senjanovic and F.~Vissani,
%``The minimal supersymmetric grand unified theory,''
arXiv:hep-ph/0306242.
%%CITATION = HEP-PH 0306242;%%
%-------------------------------------------------------------------------------------------------
%\cite{Bajc:2004xe}
\bibitem{Bajc:2004xe}
B.~Bajc, A.~Melfo, G.~Senjanovic and F.~Vissani,
%``The minimal supersymmetric grand unified theory. I: Symmetry breaking and the
%particle spectrum,''
arXiv:hep-ph/0402122.
%%CITATION = HEP-PH 0402122;%%
%-------------------------------------------------------------------------------------------------
%\cite{Bajc:2002iw}
\bibitem{Bajc:2002iw}
B.~Bajc, G.~Senjanovic and F.~Vissani,
%``b - tau unification and large atmospheric mixing: A case for non-canonical
%seesaw,''
Phys.\ Rev.\ Lett.\  {\bf 90}, 051802 (2003) [arXiv:hep-ph/0210207].
%%CITATION = HEP-PH 0210207;%%
%-------------------------------------------------------------------------------------------------
%\cite{Bajc:2001fe}
\bibitem{Bajc:2001fe} B.~Bajc, G.~Senjanovic and F.~Vissani,
%``How neutrino and charged fermion masses are connected within minimal
%supersymmetric SO(10),''
arXiv:hep-ph/0110310.
%%CITATION = HEP-PH 0110310;%%
%-------------------------------------------------------------------------------------------------
%\cite{Clark:ai}
\bibitem{Clark:ai}
T.~E.~Clark, T.~K.~Kuo and N.~Nakagawa,
%``A SO(10) Supersymmetric Grand Unified Theory,''
Phys.\ Lett.\ B {\bf 115} (1982) 26.
%%CITATION = PHLTA,B115,26;%%
%-------------------------------------------------------------------------------------------------
%\cite{Aulakh:1982sw}
\bibitem{Aulakh:1982sw}
C.~S.~Aulakh and R.~N.~Mohapatra,
%``Implications Of Supersymmetric SO(10) Grand Unification,''
Phys.\ Rev.\ D {\bf 28} (1983) 217.
%%CITATION = PHRVA,D28,217;%%
%-------------------------------------------------------------------------------------------------
%\cite{Babu:1992ia}
\bibitem{Babu:1992ia}
K.~S.~Babu and R.~N.~Mohapatra,
%``Predictive neutrino spectrum in minimal SO(10) grand unification,''
Phys.\ Rev.\ Lett.\  {\bf 70}, 2845 (1993) [arXiv:hep-ph/9209215].
%%CITATION = HEP-PH 9209215;%%
%-------------------------------------------------------------------------------------------------
%\cite{Lee:1994je}
\bibitem{Lee:1994je}
D.~G.~Lee and R.~N.~Mohapatra,
%``Automatically R conserving supersymmetric SO(10) models and mixed light Higgs
%doublets,''
Phys.\ Rev.\ D {\bf 51}, 1353 (1995)
[arXiv:hep-ph/9406328].
%%CITATION = HEP-PH 9406328;%%
%-------------------------------------------------------------------------------------------------
%\cite{Aulakh:1997fq}
\bibitem{Aulakh:1997fq}
C.~S.~Aulakh, A.~Melfo, A.~Rasin and G.~Senjanovic,
%``Supersymmetry and large scale left-right symmetry,''
Phys.\ Rev.\ D {\bf 58} (1998) 115007
[arXiv:hep-ph/9712551].
%-------------------------------------------------------------------------------------------------
%\cite{Aulakh:1999cd}
\bibitem{Aulakh:1999cd}
C.~S.~Aulakh, A.~Melfo, A.~Rasin and G.~Senjanovic,
%``See-saw and supersymmetry or exact R-parity,''
Phys.\ Lett.\ B {\bf 459} (1999) 557
[arXiv:hep-ph/9902409].
%%CITATION = HEP-PH 9902409;%%
%-------------------------------------------------------------------------------------------------
%\cite{Aulakh:2000sn}
\bibitem{Aulakh:2000sn}
C.~S.~Aulakh, B.~Bajc, A.~Melfo, A.~Rasin and G.~Senjanovic,
%``SO(10) theory of R-parity and neutrino mass,''
Nucl.\ Phys.\ B {\bf 597}, 89 (2001)
[arXiv:hep-ph/0004031].
%%CITATION = HEP-PH 0004031;%%
%-------------------------------------------------------------------------------------------------
%\cite{Bajc:2003ht}
\bibitem{Bajc:2003ht}
B.~Bajc,
%``Supersymmetric grandunification and fermion masses,''
arXiv:hep-ph/0311214 and references therein.
%-------------------------------------------------------------------------------------------------
%\cite{Goh:2003nv}
\bibitem{Goh:2003nv}
H.~S.~Goh, R.~N.~Mohapatra, S.~Nasri and S.~P.~Ng,
%``Proton decay in a minimal SUSY SO(10) model for neutrino mixings,''
Phys.\ Lett.\ B {\bf 587} (2004) 105
[arXiv:hep-ph/0311330].
%%CITATION = HEP-PH 0311330;%%
%-------------------------------------------------------------------------------------------------
%\cite{Fukuyama:2004pb}
\bibitem{Fukuyama:2004pb}
T.~Fukuyama, A.~Ilakovac, T.~Kikuchi, S.~Meljanac and N.~Okada,
%``Detailed analysis of proton decay rate in the minimal supersymmetric SO(10)
%model,''
arXiv:hep-ph/0406068.
%%CITATION = HEP-PH 0406068;%%
%-------------------------------------------------------------------------------------------------
\bibitem{general-seesaw-type-I}
M.~Gell-Mann, P.~Ramond and R.~Slansky, in {\it Supergravity}, eds. D.~Freeman et al. (North-Holland,
Amsterdam, 1980); T.~Yanagida in proc. KEK workshop, 1979 (unpublished); R.N.~Mohapatra,
G.~Senjanovi\'{c}, Phys. Rev. Lett. {\bf 44}, 912 (1980); S.L.~Glashow, {\it Cargese lectures}, (1979)
%-------------------------------------------------------------------------------------------------
\bibitem{general-seesaw-type-II}
G.~Lazarides, Q.~Shafi and C.~Wetterich,
%``Proton Lifetime And Fermion Masses In An SO(10) Model,''
Nucl.\ Phys.\ B {\bf 181} (1981) 287;
%%CITATION = NUPHA,B181,287;%%
R.~N.~Mohapatra and G.~Senjanovic,
%``Neutrino Masses And Mixings In Gauge Models With Spontaneous Parity
%Violation,''
Phys.\ Rev.\ D {\bf 23} (1981) 165;
%%CITATION = PHRVA,D23,165;%%
J.~Schechter and J.~W.~F.~Valle,
%``Neutrino Masses In SU(2) X U(1) Theories,''
Phys.\ Rev.\ D {\bf 22} (1980) 2227;
%%CITATION = PHRVA,D22,2227;%%
E.~Ma and U.~Sarkar,
%``Neutrino masses and leptogenesis with heavy Higgs triplets,''
Phys.\ Rev.\ Lett.\  {\bf 80} (1998) 5716
[arXiv:hep-ph/9802445].
%%CITATION = HEP-PH 9802445;%%
%-------------------------------------------------------------------------------------------------
%\cite{Nakaya:2002ki}
\bibitem{Nakaya:2002ki}
T.~Nakaya  [SUPER-KAMIOKANDE Collaboration],
%``Atmospheric and long baseline neutrino,''
eConf {\bf C020620} (2002) SAAT01
[arXiv:hep-ex/0209036].
%%CITATION = HEP-EX 0209036;%%
%-------------------------------------------------------------------------------------------------
\bibitem{Mohapatra1}
H.~S.~Goh, R.~N.~Mohapatra and S.~P.~Ng,
Phys.\ Lett.\ B {\bf 570}, 215 (2003) [arXiv:hep-ph/0303055].
%%CITATION = HEP-PH 0303055;%%
%-------------------------------------------------------------------------------------------------
\bibitem{Mohapatra2}
%\cite{Goh:2003hf}
H.~S.~Goh, R.~N.~Mohapatra and S.~P.~Ng,
Phys.\ Rev.\ D {\bf 68}, 115008 (2003) [arXiv:hep-ph/0308197].
%%CITATION = HEP-PH 0308197;%%
%-------------------------------------------------------------------------------------------------
%\cite{Apollonio:1999ae}
\bibitem{Apollonio:1999ae}
M.~Apollonio {\it et al.}  [CHOOZ Collaboration],
%``Limits on neutrino oscillations from the CHOOZ experiment,''
Phys.\ Lett.\ B {\bf 466} (1999) 415
[arXiv:hep-ex/9907037].
%%CITATION = HEP-EX 9907037;%%
%-------------------------------------------------------------------------------------------------
\bibitem{Eguchi:2002dm}
K.~Eguchi {\it et al.}  [KamLAND Collaboration],
Phys.\ Rev.\ Lett.\  {\bf 90} (2003) 021802
[arXiv:hep-ex/0212021].
%%CITATION = HEP-EX 0212021;%%
%-------------------------------------------------------------------------------------------------
%\cite{Ahmed:2003kj}
\bibitem{Ahmed:2003kj}
S.~N.~Ahmed {\it et al.}  [SNO Collaboration],
%``Measurement of the total active B-8 solar neutrino flux at the Sudbury
%Neutrino Observatory with enhanced neutral current sensitivity,''
Phys.\ Rev.\ Lett.\  {\bf 92} (2004) 181301
[arXiv:nucl-ex/0309004].
%%CITATION = NUCL-EX 0309004;%%
%-------------------------------------------------------------------------------------------------
\bibitem{Mohapatra3}
B. Dutta, Y. Mimura and R.~N.~Mohapatra
[arXiv:hep-ph/0402113].
%-------------------------------------------------------------------------------------------------
%\cite{Brahmachari:1997cq}
\bibitem{Brahmachari:1997cq}
B.~Brahmachari and R.~N.~Mohapatra,
%``Unified explanation of the solar and atmospheric neutrino puzzles in a
%minimal supersymmetric SO(10) model,''
Phys.\ Rev.\ D {\bf 58}, 015001 (1998)
[arXiv:hep-ph/9710371].
%%CITATION = HEP-PH 9710371;%%
%-------------------------------------------------------------------------------------------------
%\cite{Oda:1998na}
\bibitem{Oda:1998na}
K.~Y.~Oda, E.~Takasugi, M.~Tanaka and M.~Yoshimura,
%``Unified explanation of quark and lepton masses and mixings in the
%supersymmetric SO(10) model,''
Phys.\ Rev.\ D {\bf 59} (1999) 055001
[arXiv:hep-ph/9808241].
%%CITATION = HEP-PH 9808241;%%
%-------------------------------------------------------------------------------------------------
%\cite{Matsuda:2000zp}
\bibitem{Matsuda:2000zp}
K.~Matsuda, Y.~Koide and T.~Fukuyama,
%``Can the SO(10) model with two Higgs doublets reproduce the observed  fermion
%masses?,''
Phys.\ Rev.\ D {\bf 64} (2001) 053015
[arXiv:hep-ph/0010026];
%%CITATION = HEP-PH 0010026;%%
K.~Matsuda, Y.~Koide, T.~Fukuyama and H.~Nishiura,
%``How far can the SO(10) two Higgs model describe the observed neutrino  masses
%and mixings?,''
Phys.\ Rev.\ D {\bf 65}, 033008 (2002)
[Erratum-ibid.\ D {\bf 65}, 079904 (2002)]
[arXiv:hep-ph/0108202].
%%CITATION = HEP-PH 0108202;%%
%-------------------------------------------------------------------------------------------------
%\cite{Bajc:2004fj}
\bibitem{Bajc:2004fj} B.~Bajc, G.~Senjanovic and F.~Vissani,
%``Probing the nature of the seesaw in renormalizable SO(10),''
arXiv:hep-ph/0402140.
%%CITATION = HEP-PH 0402140;%%
%-------------------------------------------------------------------------------------------------
%\cite{Barbieri:1979ag}
\bibitem{Barbieri:1979ag}
R.~Barbieri, D.~V.~Nanopoulos, G.~Morchio and F.~Strocchi,
%``Neutrino Masses In Grand Unified Theories,''
Phys.\ Lett.\ B {\bf 90} (1980) 91.
%%CITATION = PHLTA,B90,91;%%
%-------------------------------------------------------------------------------------------------
%\cite{Mohapatra:1979nn}
\bibitem{Mohapatra:1979nn}
R.~N.~Mohapatra and B.~Sakita,
%``SO(2n) Grand Unification In An SU(N) Basis,''
Phys.\ Rev.\ D {\bf 21} (1980) 1062.
%%CITATION = PHRVA,D21,1062;%%
%-------------------------------------------------------------------------------------------------
%\cite{Frigerio:2002rd}
\bibitem{Frigerio:2002rd}
M.~Frigerio and A.~Y.~Smirnov,
%``Structure of neutrino mass matrix and CP violation,''
Nucl.\ Phys.\ B {\bf 640}, 233 (2002)
[arXiv:hep-ph/0202247].
%%CITATION = HEP-PH 0202247;%%
%-------------------------------------------------------------------------------------------------
%\cite{Casas:1999tg}
\bibitem{Casas:1999tg}
J.~A.~Casas, J.~R.~Espinosa, A.~Ibarra and I.~Navarro,
%``General RG equations for physical neutrino parameters and their
%phenomenological implications,''
Nucl.\ Phys.\ B {\bf 573} (2000) 652
[arXiv:hep-ph/9910420].
%%CITATION = HEP-PH 9910420;%%
%-------------------------------------------------------------------------------------------------
%\cite{Frigerio:2002in}
\bibitem{Frigerio:2002in}
M.~Frigerio and A.~Y.~Smirnov,
%``Radiative corrections to neutrino mass matrix in the standard model and
%beyond,''
JHEP {\bf 0302} (2003) 004
[arXiv:hep-ph/0212263].
%%CITATION = HEP-PH 0212263;%%
%-------------------------------------------------------------------------------------------------
%\cite{Antusch:2003kp}
\bibitem{Antusch:2003kp}
S.~Antusch, J.~Kersten, M.~Lindner and M.~Ratz,
%``Running neutrino masses, mixings and CP phases: Analytical results and
%phenomenological consequences,''
Nucl.\ Phys.\ B {\bf 674} (2003) 401
[arXiv:hep-ph/0305273].
%%CITATION = HEP-PH 0305273;%%
%-------------------------------------------------------------------------------------------------
%\cite{Melfo:2003xi}
\bibitem{Melfo:2003xi}
A.~Melfo and G.~Senjanovic,
%``Minimal supersymmetric Pati-Salam theory: Determination of physical
%scales,''
Phys.\ Rev.\ D {\bf 68} (2003) 035013
[arXiv:hep-ph/0302216].
%-------------------------------------------------------------------------------------------------
%\cite{Hagiwara:fs}
\bibitem{Hagiwara:fs}
K.~Hagiwara {\it et al.}  [Particle Data Group Collaboration],
%``Review Of Particle Physics,''
Phys.\ Rev.\ D {\bf 66} (2002) 010001.
%%CITATION = PHRVA,D66,010001;%%
%-------------------------------------------------------------------------------------------------
%\cite{Frigerio:2003zc}
%\bibitem{Frigerio:2003zc}
%M.~Frigerio,
%``Oscillations and leptogenesis: What can we learn about right-handed
%neutrinos?,''
%arXiv:hep-ph/0312023.
%%CITATION = HEP-PH 0312023;%%
%-------------------------------------------------------------------------------------------------
%\cite{Das:2000uk}
\bibitem{Das:2000uk} C.~R.~Das and M.~K.~Parida,
%``New formulas and predictions for running fermion masses at higher  scales in
%SM, 2HDM, and MSSM,''
Eur.\ Phys.\ J.\ C {\bf 20}, 121 (2001) [arXiv:hep-ph/0010004].
%%CITATION = HEP-PH 0010004;%%
%-------------------------------------------------------------------------------------------------
%\cite{Rajpoot:xy}
\bibitem{Rajpoot:xy}
S.~Rajpoot,
%``Symmetry Breaking And Intermediate Mass Scales In The SO(10) Grand Unified
%Theory,''
Phys.\ Rev.\ D {\bf 22} (1980) 2244.
%%CITATION = PHRVA,D22,2244;%%
%-------------------------------------------------------------------------------------------------
\bibitem{Mohapatraetal}
%\cite{Chang:fu}
%\bibitem{Chang:fu}
D.~Chang, R.~N.~Mohapatra and M.~K.~Parida,
%``Decoupling Parity And SU(2)-R Breaking Scales: A New Approach To Left-Right
%Symmetric Models,''
Phys.\ Rev.\ Lett.\  {\bf 52}, 1072 (1984).
%%CITATION = PRLTA,52,1072;%%
%-------------------------------------------------------------------------------------------------
%\cite{delAguila:at}
\bibitem{delAguila:at}
F.~del Aguila and L.~E.~Ibanez,
%``Higgs Bosons In SO(10) And Partial Unification,''
Nucl.\ Phys.\ B {\bf 177} (1981) 60;
%%CITATION = NUPHA,B177,60;%%
%-------------------------------------------------------------------------------------------------
%\cite{Georgi:1979md}
%\bibitem{Georgi:1979md}
H.~Georgi,
%``Towards A Grand Unified Theory Of Flavor,''
Nucl.\ Phys.\ B {\bf 156} (1979) 126.
%%CITATION = NUPHA,B156,126;%%
%-------------------------------------------------------------------------------------------------
%\cite{0406035:2004mb}
\bibitem{0406039:2004mb}
T.~Araki {\it et al.} [KamLAND Collaboration],
%``Measurement of neutrino oscillation with KamLAND: Evidence of spectral
%distortion,''
arXiv:hep-ex/0406035.
%%CITATION = HEP-EX 0406035;%%
%-------------------------------------------------------------------------------------------------
%\cite{Maltoni:2004ei}
\bibitem{Maltoni:2004ei}
M.~Maltoni, T.~Schwetz, M.~A.~Tortola and J.~W.~F.~Valle,
%``Status of global fits to neutrino oscillations,''
arXiv:hep-ph/0405172.
%%CITATION = HEP-PH 0405172;%%
%-------------------------------------------------------------------------------------------------
\end{thebibliography}
\end{document}